# Urban Green Governance: IoT-Driven Management and Enhancement of Urban Green Spaces in Campobasso

**Antonio Salis[a], Gabriele Troina[a], Gianluca Boanelli[a], Marco Ottaviano[b], Paola Fortini[b], Soraya Versace[b]**

[a] Research and Innovation Division, Tiscali Italia S.p.A., Cagliari (CA), Italy

[b] DiBT Department of Biosciences and Territory, University of Molise, Pesche (IS), Italy

**Abstract.** The efficient design and management of public green spaces is a key factor in promoting the health and well-being of urban population, as emphasized by the WHO, UNEP, and EEA. These areas serve as the "green lungs" of the urban ecosystem, playing a vital role in enhancing quality of life thanks to the provision of ecosystem services. In this context, the "Smart Green City" use case in Campobasso municipality —funded by the Italian Ministry of Enterprises and Made in Italy (MIMIT)—is an innovative model for the sustainable enhancement and management of green urban areas through the adoption of an advanced system of emerging technologies integrated and interoperable. The project integrates Internet of Things (IoT) systems and data-driven governance platforms, enabling real-time monitoring of the health status of trees and green areas via a Decision Support System (DSS). It also facilitates the collection and analysis of data from diverse sources, including weather conditions, air quality, soil moisture, pollution levels, drones, and satellites. The resulting cloud-based platform supports a holistic real time decision making for green urban managers, technical experts and operational staff, enabling the import of the urban green inventory, which is mandatory for Italian municipalities with over 15,000 inhabitants. It enables intelligent control and management of urban green spaces using "Tree Talker®" sensors, integrated with soil moisture and water potential monitoring systems. Thanks to predictive models based on machine learning algorithms and real time data provided by IoT sensors, irrigation of public parks can be optimized by providing suggestions on when and how much water to apply. Customized alerts layers are also activated warning users when monitored parameters —such as soil temperature, humidity, or water potential— exceed predefined thresholds. The Campobasso experience demonstrates how digitalization, IoT sensors fusion and technological innovation can support sustainable urban governance, fostering environmental resilience and improving citizens' quality of life.

**Keywords.** Smart Green City; Urban Governance; Technological Innovation; Distributed System; Cloud Computing; IoT; 5G; Artificial Intelligence; Tree-Talker®; Monitoring; Alerting; Dss.

# 1. Introduction

Urban green spaces are increasingly recognized as essential components of sustainable and healthy cities. Serving as the "green lungs" of urban ecosystems, these areas provide critical ecosystem services such as air purification, temperature regulation, biodiversity support, and recreational opportunities that enhance the well-being and quality of life of urban populations [1-3]. The World Health Organization emphasizes that accessible and well-managed green spaces contribute significantly to physical and mental health, reducing urban heat island effects and improving air quality, which are vital in the face of accelerating urbanization and climate change.

In parallel, the concept of smart cities has emerged as a transformative approach to managing urban growth and complexity through the integration of advanced technologies such as the Internet of Things (IoT), big data analytics, and artificial intelligence (AI). Smart cities aim to optimize resource use, improve service delivery, and enhance environmental sustainability while fostering social inclusion and economic vitality [4-5]. Within this framework, the integration of green infrastructure with smart technologies—often referred to as smart green infrastructure—represents a promising pathway to address urban environmental challenges holistically [6]. International organizations, including the United Nations and the European Environment Agency, advocate for the incorporation of nature-based solutions and green infrastructure into smart cities planning to build resilience against climate change impacts while promoting sustainable urban development [3,7].

These recommendations underscore the need for innovative management practices that leverage real-time data, sensor networks, and decision support systems to monitor and maintain urban green spaces effectively.

Recent studies highlight that embedding IoT-enabled monitoring systems within urban green areas can provide continuous, data-driven insights into tree health, soil moisture, air quality, and other environmental parameters, enabling proactive and adaptive management [8]. Such technologies facilitate optimized irrigation, pollution mitigation, and biodiversity conservation, thereby enhancing the multifunctionality of green spaces in smart cities [4].

The "Smart Green City" initiative in Campobasso exemplifies this integrated approach by deploying advanced IoT sensors, including TreeTalker® devices, coupled with cloud-based platforms and machine learning algorithms, to support real-time decision-making for urban green space management [9]. This model aligns with the growing global consensus that sustainable urban governance must harness digital innovation to foster environmental resilience and improve citizens' quality of life [10].

## 1.1 Aims and Scope

The MolisCTE project, launched under the Emerging Technologies Support Program called CTE (FSC 2014-2020), aims to foster applied research, experimentation, and technology transfer in the fields of Artificial Intelligence, IoT, and 5G, with strong involvement from local institutions and research bodies and establishment of local startups incubation and acceleration hubs. The project seeks to support the creation of startups and scalable services, accelerating the digital transformation of the territory by developing innovative digital solutions aligned with common innovation needs of local authorities. With a special focus on green ecosystems, MolisCTE promotes intelligent monitoring and management of urban green spaces, biodiversity, and air quality, leveraging smart technologies for healthier, more sustainable and efficient cities. The initiative places citizens' well-being at the center, enhancing public engagement and environmental awareness. Ultimately, MolisCTE aspires to create a replicable model for smart, eco-sustainable urban development in Campobasso and beyond [9].

The main objective of the Smart Green project in the Municipality of Campobasso is to design and pilot an innovative system for the intelligent, sustainable, and participatory management of urban green spaces, fully aligned with the latest international and European recommendations on sustainable development, biodiversity conservation, and urban resilience.

**Contest and Motivation**

Managing public green areas is a crucial challenge for local governments due to the complexity and diversity of green spaces and the growing demand from citizens for transparency, efficiency, and sustainability. In Italy, national legislation (Law 14 January 2013, no. 10) [11] and the Minimum Environmental Criteria (CAM) promoted by the Ministry for the Environment [12] emphasize the need to rationalize monitoring, maintenance, and enhancement processes for urban tree heritage, encouraging the adoption of innovative and digital solutions.

**Technological Innovation and Integrated Approach**

The Campobasso project marks a significant step forward by adopting advanced IoT technologies, particularly the Tree Talker® system [13-14]. This solution enables continuous, real-time monitoring of vital tree parameters such as growth, sap flow, wood moisture, air temperature and humidity, trunk inclination and oscillation, and solar radiation spectrum. This technology allows for the collection of large volumes of data at low cost, enabling both detailed monitoring at the individual tree level and an overall view of the city's green assets.

Integrating data collected via IoT sensors with other information sources, such as weather data, air quality, satellite and drone imagery, mobility and pollution data, enables the development of predictive models based on artificial intelligence algorithms. This optimizes maintenance planning, irrigation, risk prevention, and the enhancement of urban biodiversity.

**Operational Objectives**

The specific objectives of the project can be summarized as follows:

- **Census and digitalization of urban green assets:** To create an up-to-date and dynamic inventory of Campobasso's tree heritage and urban green, which is essential for efficient planning of maintenance activities and compliance with regulatory obligations.
- **Digitalization of urban management operation:** To create and update an inventory of urban management activities, technical tools and urban management instant messaging system of interaction among urban managers and technical staff establishing a new generation cloud-based ERP for urban green management
- **Innovative health monitoring:** To apply, for the first time in an Italian urban context, a continuous and data-driven monitoring system capable of assessing the physiological conditions of historic tree-lined avenues and quickly identifying risk or degradation situations.

- **Decision support and prevention:** To provide advanced decision support tools for public administrators and sector operators, improve operational efficiency, reduce costs, and prevent critical events (tree falls, heatwaves, extreme weather).
- **Replicable model:** To develop an integrated and intelligent urban green management model that can be adopted by other Italian and European cities, contributing to the spread of best practices and raising the standard of urban quality, with flexible operational scalability to match needs of every urban manager.
- **Expected Impact:** Through this approach, the Municipality of Campobasso aims to make the city healthier, more attractive, resilient, and livable, enhancing green spaces as strategic infrastructure for citizen well-being and urban sustainability. The project also seeks to strengthen the dialogue between the administration and the community, stimulating active participation and shared responsibility in caring for the urban environment.

### 1.2 Related Work

The contemporary discourse on urban development increasingly converges around the integrated concepts of Smart, Sustainable, and Green (SSG) cities, which collectively aim to enhance urban livability, environmental quality, and resource efficiency. Recent comprehensive reviews highlight that while each concept has distinct emphases—smart cities focusing on digital technologies and data-driven governance, sustainable cities on long-term environmental and social equity, and green cities on ecosystem services and green infrastructure—their integration is essential for addressing complex urban challenges effectively. Javidroozi et al. [15] provides a state-of-the-art synthesis emphasizing that future urban planning must adopt coordinated and integrated approaches that leverage the synergies between smart technologies and green infrastructure to foster sustainability and resilience. Their review underscores the importance of collaboration among diverse stakeholders—including government, private sector, and communities—to develop shared visions and implement multifunctional green spaces supported by smart monitoring and management systems. They also highlight the need for new financial models such as green finance and public–private partnerships to support these sustainable initiatives. At the same time urban green spaces are recognized as vital for mitigating environmental degradation caused by rapid urbanization. Studies affirm their multifunctional role in climate regulation, air quality improvement, biodiversity conservation, and enhancement of citizens' physical and mental health [16]. The expansion and effective management of green areas, supported by smart technologies, are shown to improve urban sustainability by optimizing resource use, reducing pollution, and fostering social cohesion.

The integration of IoT sensors, data analytics, and decision support systems into urban green space management represents a significant advancement in the field. These technologies enable real-time monitoring of environmental parameters such as soil moisture, air quality, and tree health, allowing for adaptive and optimized interventions like precision irrigation and pollution control [17-21]. The use of machine learning and AI further enhances predictive capabilities, supporting proactive governance and operational efficiency.

Despite progress, several challenges remain. The literature points to socio-technical barriers including data interoperability, stakeholder coordination, and the need to balance technological innovation with ecological and social considerations [15].

Furthermore, the environmental benefits of smartness are not guaranteed without explicit integration of sustainability and equity goals. Hence, future research is urged to focus on developing holistic frameworks that embed ecological principles into smart city architectures and to explore innovative governance and financing mechanisms.

European initiatives and policies increasingly promote the adoption of nature-based solutions and digital tools for urban green infrastructure enhancement, aligning with global agendas such as the UN's New Urban Agenda and Sustainable Development Goals (SDGs) [22-24]. These frameworks advocate for resilient, inclusive, and environmentally sound urban development, reinforcing the strategic role of smart green city projects as exemplified by innovative cases like Campobasso's Smart Green City initiative.

### 1.3 Outline

The paper is structured as follows: Section 2 describes the applied research method, then Section 3 describes the solution, from requirements, to the architecture to implement the scheme, then the green field put in place, and technological implementation, section 4 discusses the results, followed by a conclusion in section 5.

## 2. Research Method

The research method follows Design Science Research (DSR) [25], a well-known paradigm more oriented to design artifacts, such as the IoT system and supporting software, to solve real-world problems. This method can be split into three main cycles:

- **The Relevance cycle**, which associates the research with a real-life problem,
- **The Rigor cycle** that applies existing scientific knowledge, even integrating different sources, to contribute to new knowledge,
- **The Design cycle** that forms the actual design work the contribution of the other two cycles.

The key characteristics in this paradigm are the following:

- **Problem Oriented**: DSR works within the defined scope of the research project that requires a solution;
- **Focused on Artifacts**: DSR is centered around the definition and evaluation of artifacts, such as ideas, concepts, complex models, and related instantiations;
- **Knowledge Generation**: DSR is oriented to generate new prescriptive knowledge about how to design artifacts to solve problems in their natural setting;
- **Iterative and Evaluative**: the design and evaluation process in DSR is iterative, according to a build/measure/learn cycle. The feedback loop helps in improving the artifacts and the underlying design theory;
- **Versatility**: DSR can be applied in various disciplines and sectors, particularly in information technology and project management.

The main steps of the method are the following:

- **Problem Identification and Motivation**: Clearly define the problem and the rationale for seeking a solution, with the involvement of the final users;
- **Solution Design**: Develop a design theory, build the artifacts and implement them;
- **Design Validation**: Evaluate the artifacts in their intended context and ensure they meet the design goals;
- **Solution Implementation**: Deploy and refine the artifacts in a real-world scenario;
- **Evaluation**: Assess the artifacts' effectiveness and contribution to knowledge.

The three main cycles, as shown in Figure 1, are as follows:

- **Relevance cycle** regards all aspects related to urban green management, the emerging technologies as demanded from the CTE funded project, and business needs of the municipality of Campobasso and related management company SEA, as better described in Section 3. It includes the selection of the urban green spaces more suitable to showcase the benefits of the solution, while highlighting the resources constraints and limitation of real-time data on plant health, soil moisture, and air quality;
- **Rigor cycle** includes the application of applicable scientific knowledge on environmental and forestry aspects of plants and supporting data and technologies to contribute to novel outcomes, as well described in Section 3 and 4. It includes the definition of the IoT based system to monitor environmental parameters in real-time, optimize irrigation and maintenance, thus improving decision-making for urban planners;
- **Design cycle** covers the development of the results, getting support from the other two cycles, as depicted in Section 3 and 4. It covers the design of the IoT system in terms of sensors, IoT gateway, Cloud platform, mobile App for planners, as well as the network technology to collect all data at the edge and transfer it to the cloud. Once the prototype has been deployed, monitoring of data use the analytics to adjust the value ranges, detect and manage promptly the anomalies. Metrics ensure an objective improvement in plant health and water usage, taking into account the user feedback from city planners and operational staff.

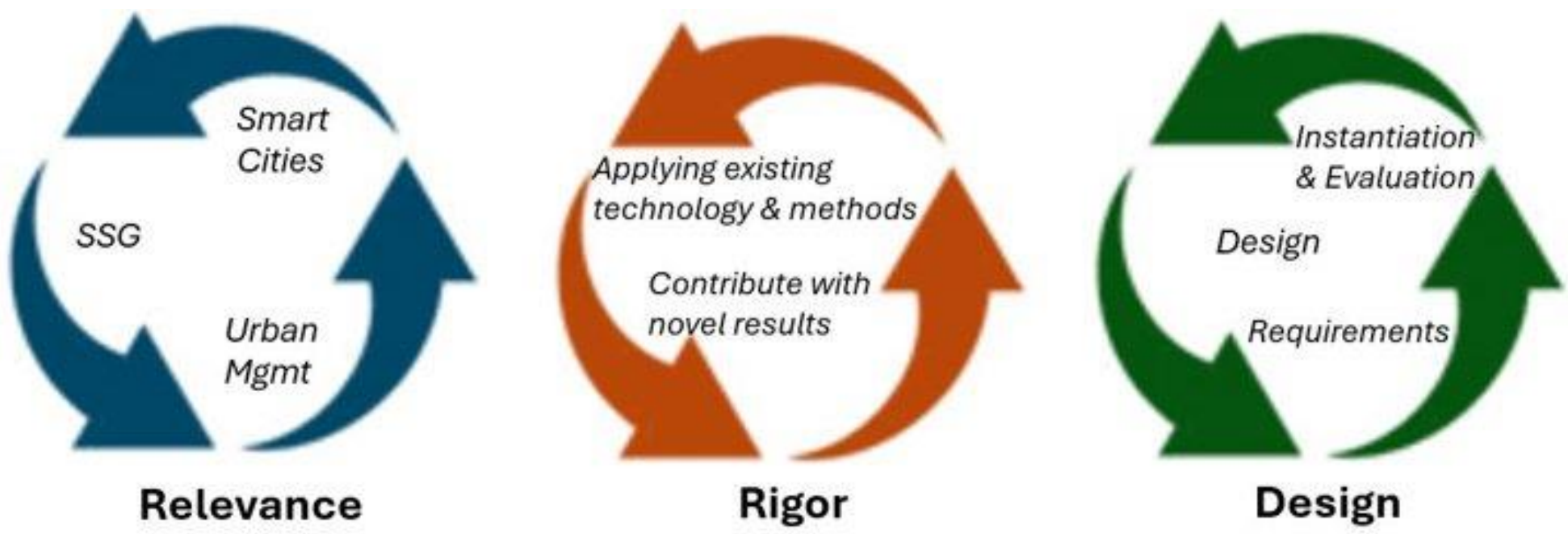

Figure 1 – A representation of the applied research methodology

## 3. The Solution

Following the guidelines suggested by the European Green Accord [26] a preliminary proposal has been outlined, with an intelligent control and management platform for urban greenery in the city of Campobasso. This proposal has been shared with experts from the Municipality of Campobasso and the green management company SEA.

The first step of problem identification has been the analysis of the area, with the classification of the types of urban green spaces present in municipal areas. All these have been mapped using GIS tools and georeferenced. Then specific urban green species have been identified, particularly oaks, which are numerous in the city. So, a decision has been made to monitor the health of these trees using innovative technologies such as micro-sensors called 'Tree Talkers', deployed in strategic areas of the municipality, integrated with soil moisture and water potential monitoring.

The envisioned system enables:

- contributing to the good health of plants and the surrounding greenery,
- collecting detailed information on plants and soils at the urban level for smart irrigation management, eliminating water waste,
- identifying critical situations for individual plants, such as the risk of trunk collapse, thus allowing timely management of potential hazards to the public,
- contributing to the conservation and protection of urban biodiversity,
- performing advanced analyses on the collected data, with results that can inform decisions on the management and maintenance of green areas, including regular pruning activities,
- monitoring air quality, helping to mitigate the negative effects of air pollution.
- Contributing to the future activities of urban area design and tree species identification

Given the distribution of relevant trees in the area, the surface area to be monitored has been shaped, a study of the sensor's installation point has been developed in order to have a regular grid of analysis. The ecosystem of resulting IoT sensors generates a real-time communication flow that leverages predictive algorithms and deep learning to generate valuable insights for the optimal management of municipal green spaces, making public green areas more accessible to citizens and significantly improving urban quality of life by prioritizing citizens' needs and well-being. The collected data can be used to create interactive maps of urban green areas and to provide useful information to citizens, such as walking routes, points of interest, and details about the species present.

To prepare for the experimentation phase, the Municipality of Campobasso has proposed several areas where Tree Talker sensors have been installed, along with weather sensors, soil moisture sensors, and air quality sensors. This setup allowed the experimentation to be carried out, the expected benefits to be assessed, and the solution to be potentially extended to the entire municipal area.

Also, in terms of Green Governance there are several benefits: the use of advanced technologies in the urban ecosystem field offers great opportunities to develop sustainable and resilient strategies for the cities of the future. Public administrations, particularly municipalities, can benefit from these technologies to achieve more efficient management of urban green areas, enabling them to:

- approve and implement Municipal Green Plans;

- develop plans to increase the quantity of green spaces in the city and improve the connectivity of green areas.

This model is replicable in any city around the world, regardless of size, and allows for data comparison between cities, helping to assess the potential impact of climate change.

**Monitored Plants**

The selection of woody plant species to be monitored using Tree Talkers has been based on several criteria. The first criterion concerns the most widespread species along the main tree-lined avenues, in historic villas, and in the most frequented public parks in the city center. A second criterion has been based on the exotic nature of many woody species that were historically used in public and private green spaces. These species are now represented by individuals that are around one hundred years old and are generally not well adapted to the city's ecological conditions, thus requiring continuous monitoring to assess their health status. Overall, 20 trees have been selected for monitoring. Fifteen species belong to the Angiosperm family and are deciduous, while five species belong to the Gymnosperm family and are evergreen. Two of the selected species are listed on the IUCN Red List: *Sequoia sempervirens*, classified as EN (Endangered), and *Aesculus hippocastanum*, classified as VU (Vulnerable), both with a global population trend described as “decreasing”.

Thus, the urban areas of Campobasso selected for the installation of TreeTalker have been Villa De Capoa, Villa Musenga, and Piazza Vittorio Emanuele II, where the selected species are present, as shown in Figure 2.

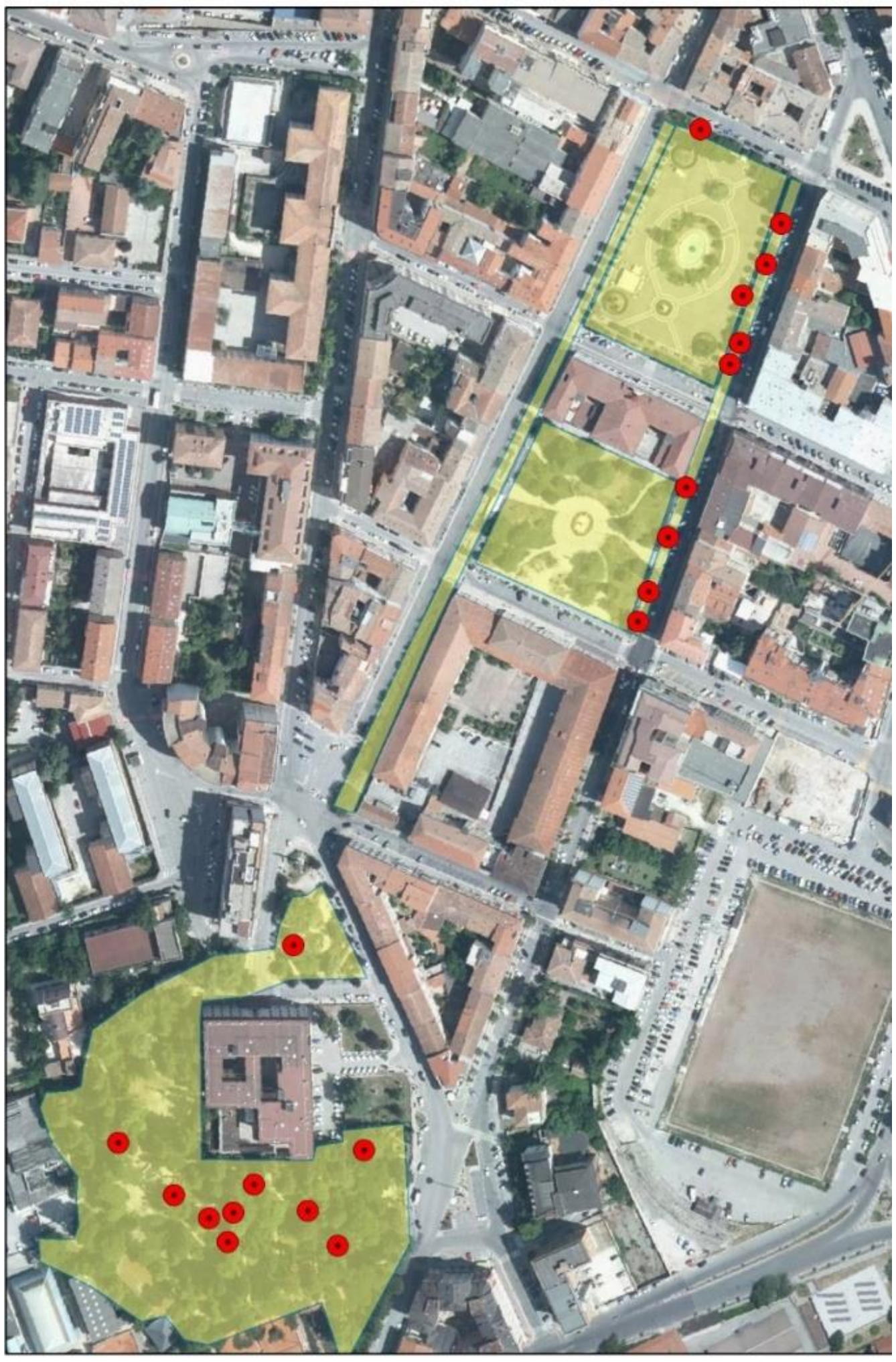
Figure 2 – The areas selected for the installation of TreeTalker and sensors, red spots representing the trees to be monitored

**Sensors**

Once the urban areas and plants to be monitored have been defined, the set of rechargeable sensors to be installed has been defined, with related positions. It includes the following:

- **2 weather stations** with solar radiation sensor, able to collect the following parameters every 30 minutes: dry air temperature, wet bulb temperature, dew point, air humidity, wind speed and direction, cumulative rainfall, leaf wetness and humidity, solar radiation;
- **2 air quality sensors** able to collect the following parameters every 30 minutes: air temperature and humidity, $CO_2$, Particulate matter: PM 1 – 2.5 – 4 – 10;

- **4 electro sensors** able to collect the following parameters every 30 minutes: soil moisture, temperature and salinity;
- **4 hydro sensors** able to collect the following parameters every 30 minutes: water potential and soil temperature;
- **20 Tree Talkers (TT),** a non-invasive technology that integrates various sensors to measure both physiological variables of individual trees and ecological-environmental variables influenced by the presence of the tree itself. The device consists of a main body equipped with two sensors that are applied to the tree bark, able to measure water transport within the plant and its radial growth. Other variables such as **Foliage Health** (light transmission measured in 12 spectral bands), **tree stability** (gyroscopic sensor), and **air and soil temperature and humidity** are measured by external sensors integrated into the device. Each tree involved in the experiment has been georeferenced, coded, and classified.
- **Remote sensing Satellite data:** Sentinel Satellites are part of the ESA Copernicus programme. Sentinel-2 has a High-Resolution Multispectral Optical Sensor, with a spatial resolution of 10 meters for RGB and Near-Infrared bands, and a revisit frequency of 5 days under cloud-free conditions. They are ideal for crop monitoring, vegetation indices, forest monitoring, land-use mapping, water resource management, climate change monitoring.
- **UAV (Unmanned Aerial Vehicle) sensors**, DJI Mavic Enterprise multispectral and DJI Matrice 350 with LiDAR L2, with Horizontal Accuracy of 5cm @150m, Vertical Accuracy of 4cm @150m, Ranging Accuracy of 2cm @150m, which, through targeted campaigns, made it possible to acquire information on the dendro-structural and physiological characteristics present in the study areas. The ground-acquired data is thus integrated with multitemporal data collected by drones equipped with multispectral and LiDAR sensors, allowing the processing of true-color images and indices related to physiological activity.
- **3 LoraWAN Gateways,** to collect all data coming from sensors and transfer them to the server hosted in a cloud environment via a 5G device for appropriate data analysis. The resulting system architecture is shown in Figure 3.

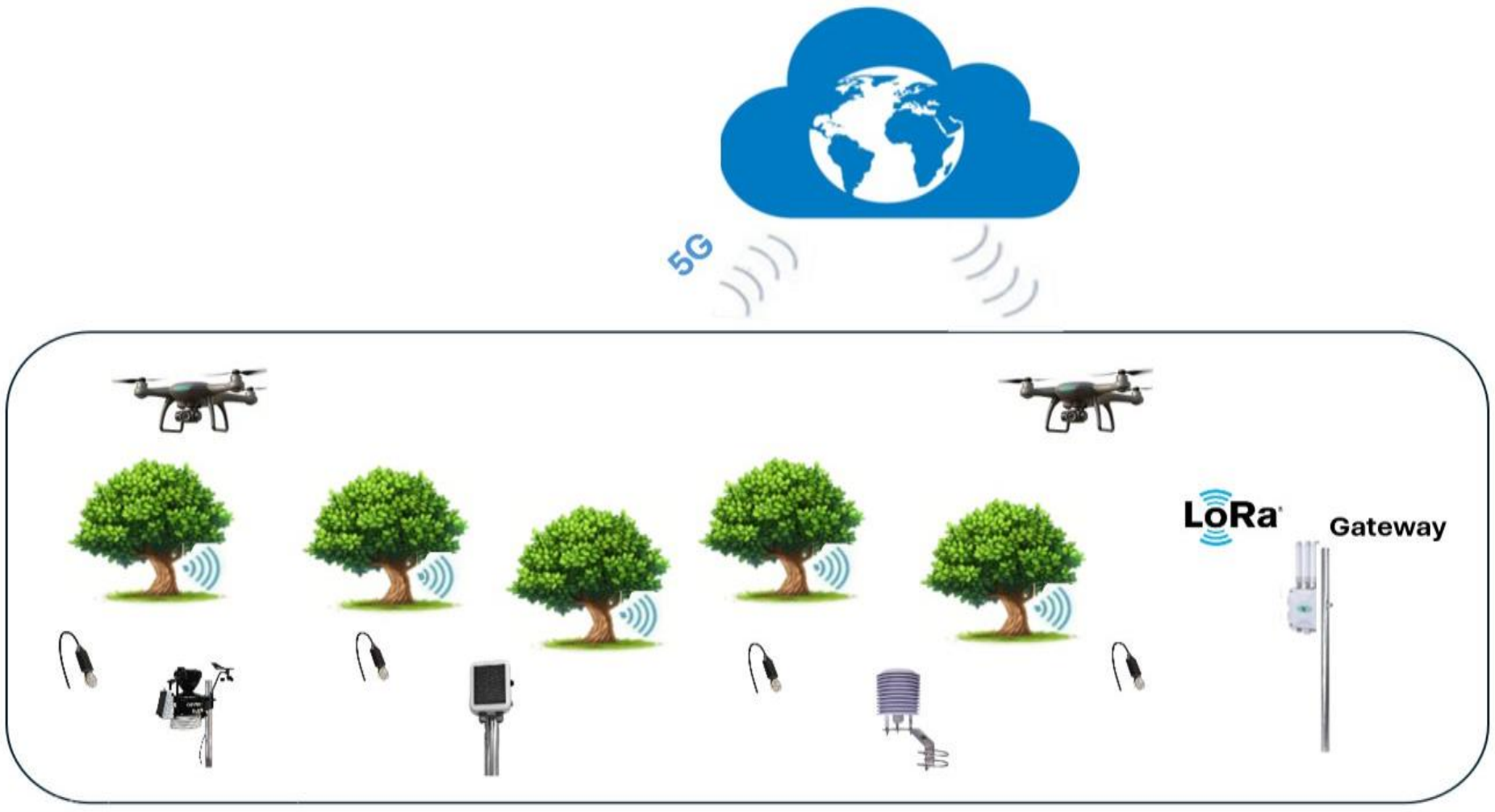

Figure 3 – Distributed System Architecture

**Data Processing Workflow**

The following outlines the key steps in creating and utilizing the urban green monitoring system:

1. Installation of Tree Talkers and the TT-Cloud concentrator, including LoRaWAN coverage verification. This involves setting up 5G connectivity from the TT-Cloud to the Cloud. This step enables the **Download Spatial Data** phase.
2. The Cloud hosts the server and storage for data archiving. To enhance the usability of the final tool, some local information will be imported. This includes municipal satellite imagery, supplemented by high-resolution drone scans of specific areas, along with any relevant weather data and information on forest species present in the experimental zones. It enables also the import of the urban green inventory. This step enables the **Import Spatial Data** phase.
3. On the Cloud, data undergoes pre-processing to eliminate spurious data (cleaning), and appropriate data mappings are performed for subsequent analysis. This step enables the **Pre-Process Spatial Data** phase.
4. The data is then loaded onto the GIS system, making it available for visualization by the administrator. This step enables the **Various Tasks** phase.
5. Data analysis, including predictive algorithms and deep learning, is performed on the Cloud. The output of this analysis can be represented as layers to be displayed on the map. This step enables the **Analyze Data** phase.
6. Finally, the administrator can visualize the processing results on interactive maps. This step enables the **Create Maps** phase.

**The Verdeview solution**

All data collected by the sensors is gathered and analyzed on the "Verdeview" platform, which provides decision support and is designed to meet the needs of the urban green space management company of Campobasso. The platform is hosted on the AWS (Amazon Web Services), within an instance located in the European region. The servers are regularly updated with operating system upgrades and supporting tools, and they feature autoscaling to guarantee efficient management of traffic and ensure high service quality. Data in transit is encrypted using industry-standard encryption protocols, and all resources use HTTPS/TLS via AWS Certificate Manager.

The Manager, that is, the staff of the Municipality of Campobasso, as a first step, graphically identifies the urban areas of interest for monitoring by importing existing layout files or drawing them manually, after locating the areas of interest using GPS data.

The 'Weather' module allows access to meteorological data provided by specific weather stations distributed throughout the testbed areas. This enables the manager to use forecasts to effectively plan work and to leverage the data to optimize the management of urban green spaces.

Historical data series are also available for up to 90 days prior to the current date, allowing the manager to monitor changes in environmental conditions over time and compare data collected during different periods. This also helps identify any trends or anomalies that may affect the management and maintenance of green areas.

The ‘Tasks’ module allows the manager to use the platform as a tool for planning, scheduling, and reporting maintenance activities for urban green spaces, such as pruning, as shown in Figure4. It involves relevant personnel as users, including teams responsible for routine and extraordinary maintenance operations.

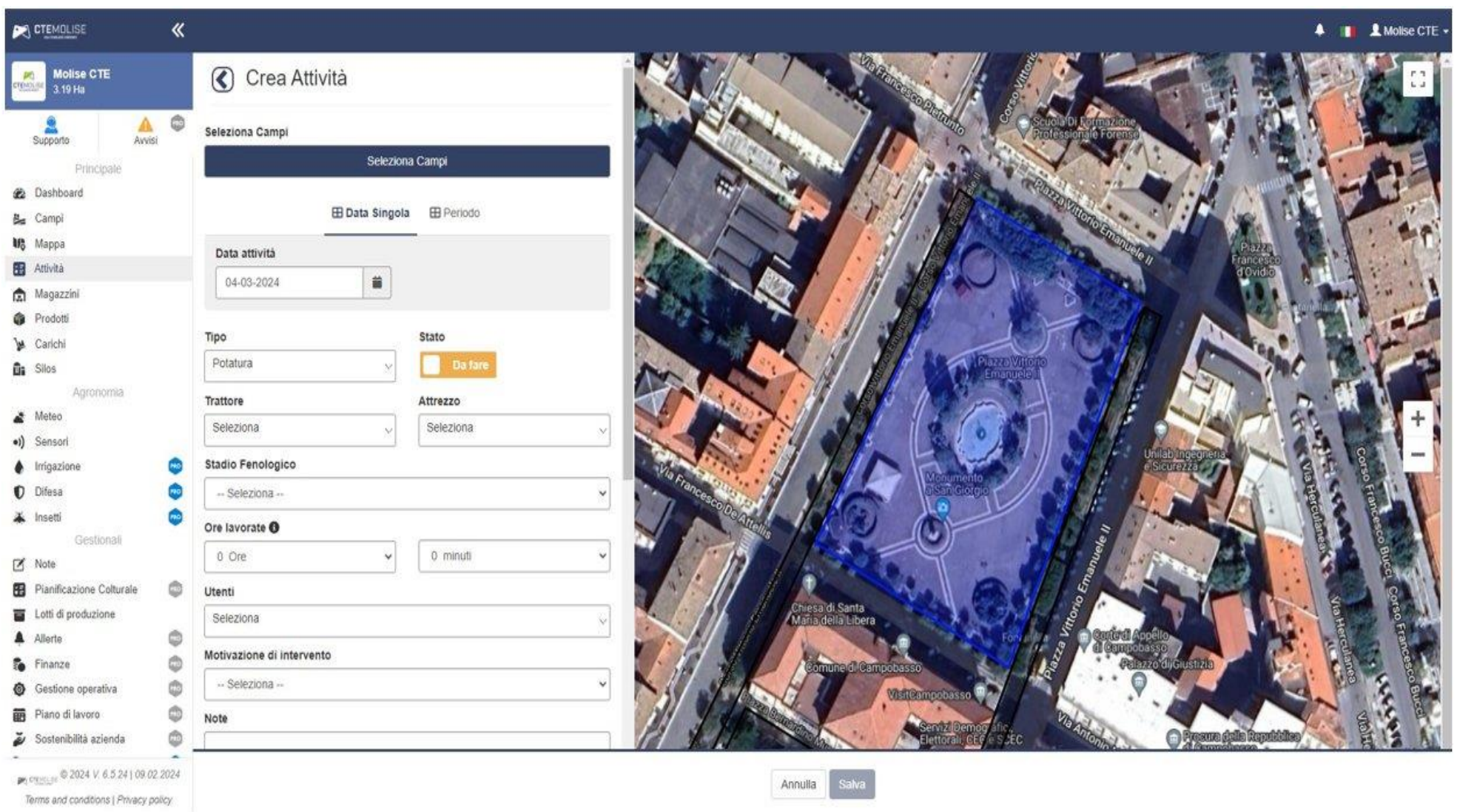

Figure 4 – A screenshot showing the planning of pruning of an oak

The 'Controls' module enables the configuration of custom alerts to allow notifications to be sent (via SMS, email, and in-app) to technical experts and operational staff when monitored levels, such as temperature, humidity, soil water potential, exceed the configured thresholds.

The 'Maps' module enables the manager to quickly identify the geographic coordinates of all installed sensors, thanks to GPS positioning.

The 'Satellite' module allows monitoring of various vegetation indices over the areas of interest using the latest multispectral satellite image, updated every 3 to 5 days depending on cloud cover, or using data from drone-based remote sensing. For example, with the NDVI (Normalized Difference Vegetation Index), as shown in Figure 5, it is possible to monitor vegetation growth and vigor in different areas through changes in the color scale.

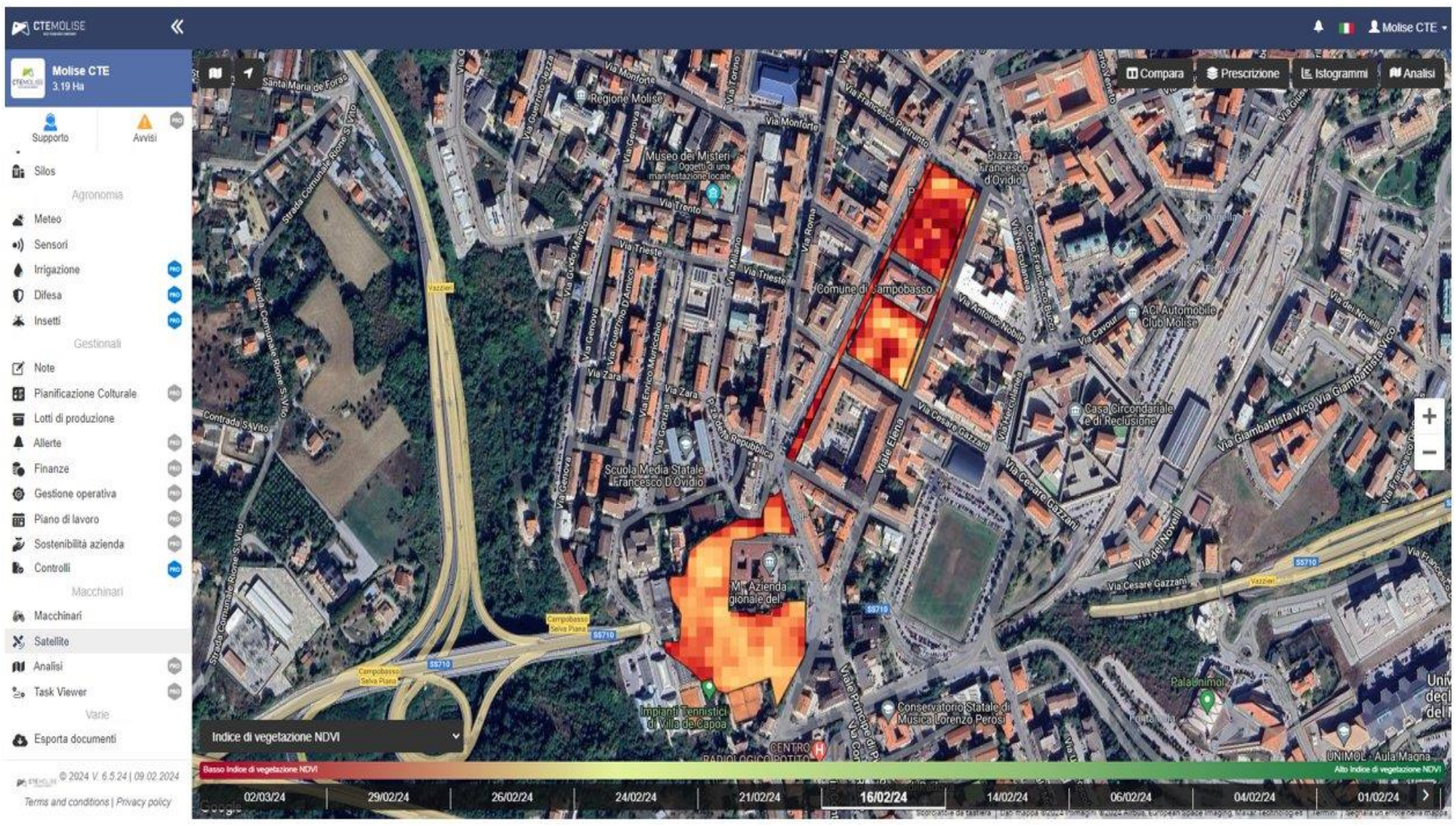

Figure 5 – A screenshot showing the NDVI analysis in the test field

## 4. Results

The validation phase of the VerdeView platform, conducted outdoors in the municipality of Campobasso, made it possible to systematically test the effectiveness of the management modules and the IoT sensors installed in some strategic urban areas.

Specifically, over a year of monitoring, air and soil temperatures were recorded in degrees Celsius. During the monitoring period, from June 27, 2024, to June 27, 2025, air temperature showed a regular seasonal trend, with higher values between July and mid-September, often exceeding 30°C, and a gradual decrease starting in October. The coldest months were December, January, and February, with temperatures often below 5°C and occasional drops close to zero.

Similarly, the soil temperature graph shows a comparable seasonal trend. The surface soil temperature, measured by two probes, reached maximum values above 25°C during the summer months and gradually decreased to around 10–12°C between November and December, remaining more stable and with less thermal fluctuation in winter compared to air temperature. From March onward, a new temperature rise was observed, with values exceeding 20°C by the end of June.

Regarding the thermal trends recorded during the monitoring year, the NDVI vegetation index showed fluctuations between -0.3 and 0.7 from January to June 2025. In the winter months, values near or below zero were consistent with the low air and soil temperatures recorded during the same period, conditions that resulted in reduced photosynthetic activity and virtually no vegetative growth. The beginning of spring, from late March, coincided with a gradual temperature rise, which was also reflected in the NDVI values. This index showed a gradual increase, becoming more pronounced between April and May, reaching values above 0.5 between May and June, indicating active leaf cover recovery and good physiological condition of the tree vegetation.

Trunk growth data (measured in millimeters) also showed a seasonal trend closely linked to temperature and the vegetative cycle. In the colder months up to mid-April, diameter growth values remained low, between 0 and 2.5 mm, indicating the physiological phase of vegetative dormancy. From late spring, with rising temperatures and improved environmental conditions, a significant growth increase was observed, with values exceeding 4 mm between May and June, a period characterized by more intense physiological and growth activity.

The ground-acquired data, integrated with multitemporal data collected by drones equipped with multispectral and LiDAR sensors, has allowed the processing of true-color images and indices related to physiological activity, as shown in Figure 6, such as:

- NDVI (Normalized Difference Vegetation Index)
- LCI (Leaf Chlorophyll Index)
- CHM (Canopy Height Model)

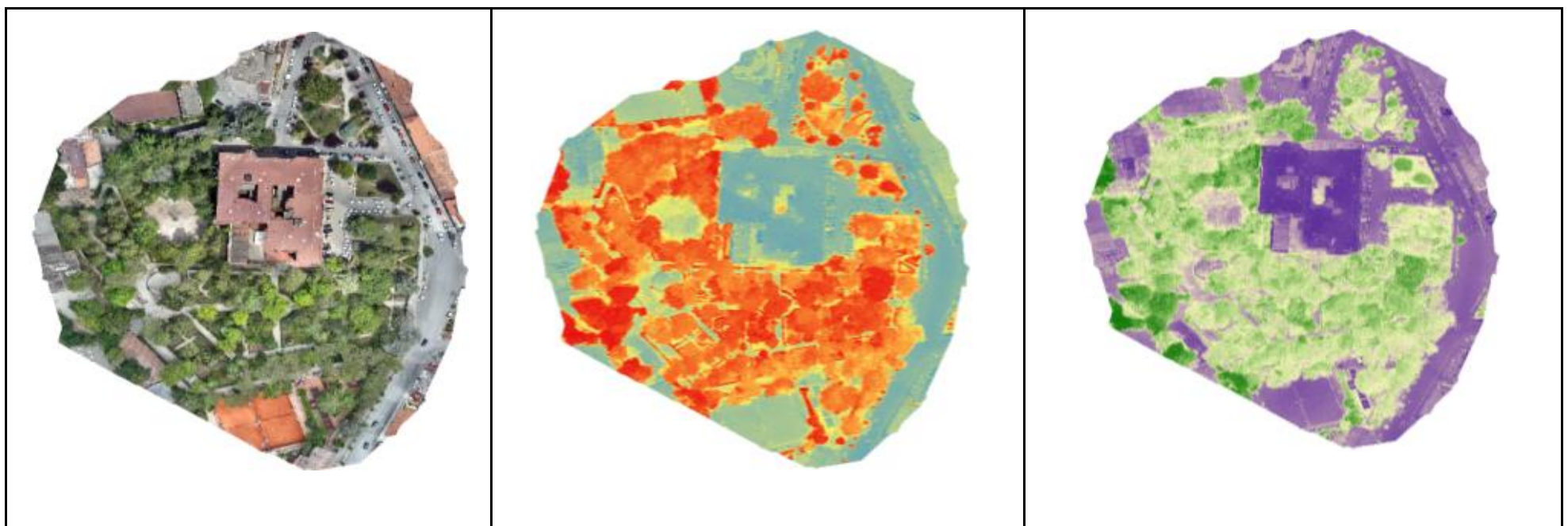

| *Composite in real colors* | *NDVI (Normalized Difference Vegetation Index)* | *LCI (Leaf Chlorophyll Index)* |
|---|---|---|

Figure 6 – Picture of Villa De Capoa according with different vegetation indexes

The results demonstrated the platform's ability to support environmental monitoring and the operational management of urban greenery. In particular, the analysis of soil salinity data showed average values between 0.3 and 0.6 dS/m, with anomalous peaks near areas subject to waterlogging, suggesting the need for targeted corrective actions. For this, specific alerts have been configured driving the prompt resolution of the issue. The water balance, monitored through soil moisture and water potential sensors, revealed discrepancies between the recorded data and actual conditions, as in the case of a rain gauge obstructed by foliage, which initially registered zero rainfall during significant rain events. This episode highlighted the importance of data quality control and led to an improvement in sensor placement. Another aspect that emerged soon has been the correct monitoring of the battery load of sensors; solar panels have been moved from the original placement to guarantee the necessary energy recharge. Apart from this case all sensors do not require frequent maintenance. The sensors played a crucial role also in monitoring tree stability, successfully detecting abnormal movements in one case. This issue was effectively resolved through a targeted pruning intervention.

Therefore, we can conclude that the adopted technologies enabled the collection, processing, and integrated analysis of heterogeneous data, including air and soil temperature, trunk growth, and physiological responses of trees, complemented by drone surveys and satellite data. This information is particularly useful for managing urban greenery, as it allows for a deeper understanding of the relationships between environmental conditions and plant physiological dynamics. In particular, soil temperature directly influences fundamental processes such as root absorption, respiration, and microbial activity, while air temperature affects photosynthesis, transpiration, and the plant's water balance. Integrating this environmental data into the Decision Support System (DSS) allows for targeted and timely planning of irrigation, pruning, and phytosanitary management operations, optimizing time, resources, and interventions based on real-time conditions.

### 4.1 Assessment of Results

The innovative aspect of the experiment lies in the integration of physiological plant data (sap flow, diameter growth, stability) with environmental and management parameters, providing a holistic and dynamic view of the health status of urban greenery. The project's originality is reflected in its ability to generate personalized alerts and automated intervention plans, while its significance is demonstrated by its capacity to concretely address the needs of the municipality of Campobasso in terms of sustainability, operational efficiency, and enhancement of green assets. The results obtained contribute to the state of the art in the field of intelligent urban monitoring, proposing a replicable model for other municipalities.

The validation process was guided by the Design Science Research (DSR) methodology, which provided a structured framework for iteratively refining the technological artefact—the VerdeView platform—based on real-world feedback and performance data. DSR emphasizes the creation and evaluation of innovative solutions to complex problems through rigorous testing and contextual adaptation. In this case, the methodology enabled the identification of critical operational issues (e.g., sensor misplacement, data anomalies, energy inefficiencies) and facilitated their resolution through targeted design interventions.

The benefits of applying DSR in this context are multifold:

- **Enhanced reliability and robustness** of the platform through iterative validation and refinement cycles.
- **Improved alignment with stakeholder needs**, ensuring that the system evolves in response to actual urban management challenges.
- **Scalable insights** for future deployments, as the methodology supports generalization of findings across similar urban contexts.

- **Evidence-based innovation**, where each design decision is grounded in empirical data and contributes to the theoretical advancement of smart urban ecosystem management.

By embedding DSR into the validation phase, the VerdeView project not only demonstrated technical feasibility but also laid the groundwork for a replicable and adaptable model of smart green infrastructure management.

### 4.2 Limitations

The required data exhibit seasonal patterns, making it essential to collect information over multiple years to effectively train machine learning (ML) algorithms. Given the 24-month duration of the project, the dataset gathered was sufficient to construct a time series suitable for the intended ML applications. The integration of high-resolution data from drone scans significantly enhanced the dataset's richness, thereby improving the accuracy of predictive models. Nevertheless, extending the time series would provide a more robust foundation for the development and optimization of AI algorithms, as well as chance to optimize frequency of transmission of sensors data, thus extending the life of batteries.

### 4.3 Further Research

Looking ahead, future research should aim to expand the coverage of green areas within the municipality. This would entail the management of a broader range of arboreal species, accompanied by targeted operational treatments and continuous monitoring. Simultaneously, increasing the number, variety, and density of sensors would enrich the dataset, enabling the development and refinement of advanced predictive models to enable Smart Irrigation and Decision Support System targeting specifically arboreal species. The resulting platform could then be adapted and implemented in other municipalities, both within Italy and internationally, to capitalize on the insights gained. In this context, MIMIT actively promotes cross-pollination among different CTEs projects across cities, fostering the scalability and replication of successful initiatives throughout the Italian territory.

## 5. Conclusion

The Smart Green City project in Campobasso has achieved its primary objectives, demonstrating the effectiveness of integrating IoT technologies, predictive analytics, and cloud-based platforms for sustainable urban green governance. During the project's lifecycle, municipal experts and managers consistently provided positive feedback, emphasizing the platform's intuitive interface, the reliability of real-time monitoring, and the significant improvements in operational efficiency and decision-making for urban green space management.

The research question—whether advanced digital tools and IoT systems can enhance the management and sustainability of urban green areas—has been positively answered. The experimental phase produced clear evidence of improved plant health, optimized irrigation, real time defense action to phatogens**,** and more effective planning and maintenance activities, confirming the model's replicability for other municipalities.

Furthermore, the project has received appreciation from MIMIT, who acknowledged the innovative approach and the project's alignment with national strategies for digital transformation and environmental sustainability. The Minister highlighted the initiative as an exemplary model for smart, citizen-centered, and eco-sustainable urban development, reinforcing its strategic importance for Italy's future.

In summary, the Smart Green City project has set a new standard for urban green governance, combining technological innovation with a strong focus on citizen well-being and institutional collaboration. The positive feedback from local stakeholders and the recognition by national authorities underscore the project's success and its potential for broader adoption.

**Acknowledgments** This work is supported by the Molise CTE Project, funded by MIMIT (FSC 2014- 2020), grant #D33B22000060001.